# Continuous isotropic – nematic transition in compressed rod-line liquid crystal based nanocolloid

Joanna Łoś[1]*, Aleksandra Drozd-Rzoska[1]*, Sylwester J. Rzoska[1], Szymon Starzonek[1]

Krzysztof Czupryński[2], Prabir Mukherjee[3]

[1]Institute of High Pressure Physics Polish Academy of Sciences,
ul. Sokołowska 29/37, 01-142 Warsaw, Poland

[2]Military University of Technology, Faculty of Advanced Technologies and Chemistry,
ul. gen. Sylwestra Kaliskiego 2, 00-908 Warsaw, Poland

[3]Department of Physics, Government College of Engineering and Textile Technology,
12 William Carey Road, Serampore, Hooghly-712201, India

Joanna Łoś:
ORCID: 0000-0001-6283-0948; e-mail: joalos@unipress.waw.pl
Aleksandra Drozd-Rzoska:
ORCID: 0000-0001-8510-2388;
e-mail: Ola.DrozdRzoska@gmail.com
Sylwester J. Rzoska:
ORCID: 0000-0002-2736-2891; e-mail: sylwester.rzoska@gmail.com
Szymon Starzonek:
ORCID: 0000-0003-2793-7971; e-mail: starzonek@icloud.com
Krzysztof Czupryński:
ORCID: 0000-0002-1785-7786; e-mail: krzysztof.czuprynski@wat.edu.pl
Prabir K. Mukherjee:
ORCID: 0000-0003-4523-5391 ; e-mail: pkmuk1966@gmail.com

**Abstract**

Landau-de Gennes mean-field model predicts the discontinuous transition for the isotropic – nematic (I-N) transition, associated with uniaxial ordering and a quadrupolar order parameter in three dimensions. This report shows pressure-related dielectric studies for rod-like nematogenic pentylcyanobiphenyl (5CB) and its nanocolloids with $BaTiO_3$ nanoparticles. The scan of dielectric constant revealed the continuous I-N phase transition in the compressed nanocolloid with a tiny amount of nanoparticles ($x = 0.1\%$).

For the nematic phase in 5CB and its $x=1\%$ nanocolloid the enormous values of dielectric constant and the bending–type, long-range pretransitional behavior were detected. The 'shaping' influence of prenematic fluctuations was also noted for the ionic-related contribution to dielectric permittivity in the isotropic liquid phase. For the high-frequency relaxation domain, this impact was tested for the primary relaxation time pressure evolution and the translational–orientational decoupling.

**Introduction**

The phase transition from isotropic liquid (I) to nematic phase (N) in rod-like liquid crystalline materials can be considered a unique case of challenging fluid-fluid transition phenomenon [1-3] due to the fluidity of neighboring phases. It is also an example of a freezing/melting transition associated with only a single element of symmetry: the uniaxial arrangement. I-N is also the entrance transition to the 'intermediate' state of matter, liquid crystals (LC) which combine fluidity with a limited crystalline ordering. Notable is the 'weakness' of the I-N phase transition [4-6]. For instance, in such classic LC material as pentylcyanobiphenyl (5CB) the entropy changes at the I-N transition is $\Delta S \approx 8\, JK^{-1}kg^{-1}$ for the I-N transition, whereas $\Delta S \approx 160\, JK^{-1}kg^{-1}$ for the nematic – solid (N-S) transition [7, 8]. The hallmark feature of the I-N transition is long-range pretransitional effects associated with critical-like prenematic fluctuations in the isotropic liquid phase [4-6].

Five decades ago, Pierre Gilles de Gennes considered the simple 'universal; temperature changes of the Cotton Mouton effect ($CME$, $\Delta n/\lambda H^2$) and Rayleigh light scattering ($I_L$) in the isotropic liquid phase [9, 10]. Later similar behavior was also observed for the Kerr effect ($KE$, $\Delta n/\lambda E^2$), nonlinear dielectric effect ($NDE$, $\Delta\varepsilon/E^2$) and compressibility ($\chi_T$). All these effects follow the same simple scaling pattern [4, 6, 9-20]:

$$\frac{\Delta n}{\lambda H^2}, \frac{\Delta n}{\lambda E^2}, \frac{\Delta\varepsilon}{E^2}, I_L, \chi_T = \frac{m}{T-T^*} \qquad (1a)$$

$$\left(\frac{\Delta n}{\lambda H^2}\right)^{-1}, \left(\frac{\Delta n}{\lambda E^2}\right)^{-1}, \left(\frac{\Delta\varepsilon}{E^2}\right)^{-1}, (I_L)^{-1}, \chi_T^{-1} = mT - mT^* \qquad (1b)$$

where *m* denotes the amplitude related to the given method, $\Delta n$ is the magnetic (*H*) or electric (*E*) fields induced birefringence, $\lambda$ denotes the applied light wavelength, $\Delta\varepsilon = \varepsilon - \varepsilon(E)$ is for electric field induced changes of dielectric constant, $T > T_{I-N}$ and $T^* = T_{I-N} - \Delta T^*$, the latter is the temperature metric for the I-N transition discontinuity.

In refs. [9, 10] de Gennes linked Landau model associating pretransitional behavior for the continuous phase transition with local order parameter changes [21] and the nematic order parameter definition introduced by Tsvetkov [22, 23]. It led to the following free energy expansion [9, 10, 24]:

$$F = F_0 + aQ_{\alpha\beta}Q_{\beta\alpha} - bQ_{\alpha\beta}Q_{\beta\gamma}Q_{\gamma\alpha} + c(Q_{\alpha\beta}Q_{\beta\alpha})^2 - G(H_\alpha H_\beta, E_\alpha E_\beta) \qquad (2)$$

where $Q_{\alpha\beta} = \frac{1}{2}S(3 n_\alpha n_\beta - \delta_{\alpha\beta})$ is the quadrupolar order parameter reflecting the equivalence of $\vec{n}$ and $-\vec{n}$ directors indicating the preferred uniaxial ordering of rod-like molecules [22, 23]; *b, c, G* are constant amplitudes [9, 10, 24]. In the isotropic phase: $a = a_0(T - T^*)$. The last term inEq. (2) reflects the impact of the magnetic or electric field, respectively.

Substituting the above of order parameter to Eq. (2) one obtains the simplified relation, with the scalar order parameter metric [4, 6, 24]:

$$F = F_0 + \frac{3}{2}aS^2 - \frac{3}{4}bS^3 + \frac{9}{4}cS^4 - \cdots G(H^2, E^2) \qquad (3)$$

Recalling the *Physics of Critical Phenomena* [4] for which critical exponents are essential universal parameters, the basic Landau – de Gennes (LdG) model yields: $\gamma = 1$ for the compressibility-related exponent, as in Eq. (1), and $\beta = 1/2$ for the order parameter exponent in the nematic phase. For such basic magnitude as the specific heat exponents $\alpha = 1/2$ in the nematic phase and $\alpha = 0$ in the isotropic liquid phase in the mean field approximation is expected [4, 25-28]. The latter is in explicit disagreement with extensive experimental evidence showing $\alpha = 1/2$ for both sides of $T_{I-N}$ [4, 6, 29]. Such values of the specific heat exponent can be obtained assuming that I-N transition occurs in the vicinity of the tricritical point, also associated with $\gamma = 1$ and $\beta = 1/4$ [4, 30]. Notwithstanding the exponent $\alpha = 1/2$ can also appear in the basic mean field LdG model when considering corrections with the density changes impact associated with prenematic fluctuations [31]. There is some split in the evidence of the order parameter exponent in the nematic phase. There is a set of reliable experimental results yielding $\beta = 1/2$ in agreement with the basic mean-fied LdG model approach [4, 24, 32-34]. On the other hand, the distortions sensitive analysis reduced the number of parameters in heptyloxycyanobiphenyl (6OCB) [35], octyloxycyanobiphenyl (8OCB) [36], and 5CB [37] yielded $\beta = 1/4$.

In liquids, in the vicinity of a continuous phase transition, changes of specific heat are coupled to the pretransitional behavior of dielectric constant, $c_v(T) = d\varepsilon(T)/dT$ [38, 39]. This theoretical conclusion confirms the extensive experimental evidence in the isotropic phase of nematogenic LC compounds [36, 37, 40, 41]:

$$\varepsilon(T) = \varepsilon^* + a_\varepsilon(T - T^*) + A_\varepsilon(T - T^*)^{1-\alpha} \qquad (4)$$

The significance of dielectric constant as the material characterization is notable with respect to the fundamental insight [4, 6, 32] and applications. The latter is related to its role in indicating the ability to interact with the external electric field, which is essential in omnipotent devices using LC rod-like materials [32, 42].

When discussing I-N transition, three important fundamental challenges have to be indicated.

First, Landau-de Gennes mean-field model predicts the discontinuous transition for the isotropic – nematic (I-N) transition, associated with uniaxial ordering and a quadrupolar order parameter in three dimensions. Consequently, there is a common consensus for the generic discontinuity for I-N transition in rod-like systems [4, 6, 24, 29, 32]. We omit here the quasi-continuous case when the phase transition is stretched in some temperature domain [43-46].

Second, rod-like LC systems can serve as a unique 'experimental model' facilitating the theoretical and simulation insights into the still challenging glassy dynamics domain [47-51].

Third, LC systems are enormously sensitive to endogenic and exogenic impacts, such as pressure [17, 37] and the addition of nanoparticles (NPs) [50-53], for instance

The latter example recalls the great area of LC-based nanocolloids and nanocomposites enormously developing in the last decades It is related to emerging unique fundamental properties of such systems, but the key motivator constitutes possibilities of applications in innovative photonic devices [52-54].

However, there are significant cognitive gaps in studies on LC+NPs nanocolloids, namely: (i) the limited evidence regarding the impact of pretransitional fluctuations essential in basic LC compounds, and (ii) the practical lack of pressure-related studies in LC-based nanocolloids.

Notable that LC mesophases can be alternatively created by cooling and compressing ]45]. However, cooling changes the activation energy [55, 56], whereas compressing the activation volume [57, 58]. It opens significant cognitive possibilities [55].

This report presents high-pressure studies of 5CB and its nanocolloids with $BaTiO_3$ nanoparticles, focused on the impact of pretransitional fluctuations and the emergence of complex, glassy-type dynamics.

Amongst the results obtained worth stressing is the unique evidence for the continuous I-N transition emerging in compressed 5CB+NPs nanocolloid.

**Experimental**

Studies were carried out in 4-pentyl-4'-cyanobiphenyl (pentylcyanobiphenyl, 5CB), the rod-like liquid crystalline (LC) compound, with the following mesomorphism [32]:

$Crystal \leftarrow 288.0\,K \leftarrow Nematic \leftarrow 308.2\,K \leftarrow Isotropic$. 5CB belongs to the most classical LC compounds regarding applications and fundamental research, where it reached the 'experimental model system' status. 5CB molecule is approximately rod-like and has a permanent dipole moment approximately parallel to the long molecular axis: $\mu = 5.95\,D$ [32] The compound was synthesized at the Military University of Technology (Poland), in Krzysztof Czuprynski team. The sample was deeply cleaned to reduce contaminations, which can yield a parasitic rise in electric conductivity. It was also degassed immediately prior to measurements. Paraelectric $BaTiO_3$ nanopowder (diameter $d = 50$ nm) was purchased from US Research

Nanomaterials, Inc. [59]. Mixtures of liquid crystal and nanoparticles were sonicated at a temperature higher than the isotropic to nematic phase transition for 4 hours to obtain homogeneous suspensions. Studies were carried out for 5CB+0.1% and 5CB +1% of $BaTiO_3$ nanoparticles. Paraelectric properties and the globular form of nanoparticles (NPs) enabled to minimization of their phase- and shape-related orientational impact on the host LC system. Concentrations of nanoparticles are given in weight fraction percentage (*wt%*). Tested samples were placed in a flat-parallel capacitor (diameter *2r* = 15 mm), made of Invar and with the distance between plates *d* = 0.15 mm. The capacitor was placed in the high-pressure chamber (UnipressEquipment). The sample was perfectly isolated from the medium (Plexol) transmitted pressure due to the design of the measurement capacitor [60].

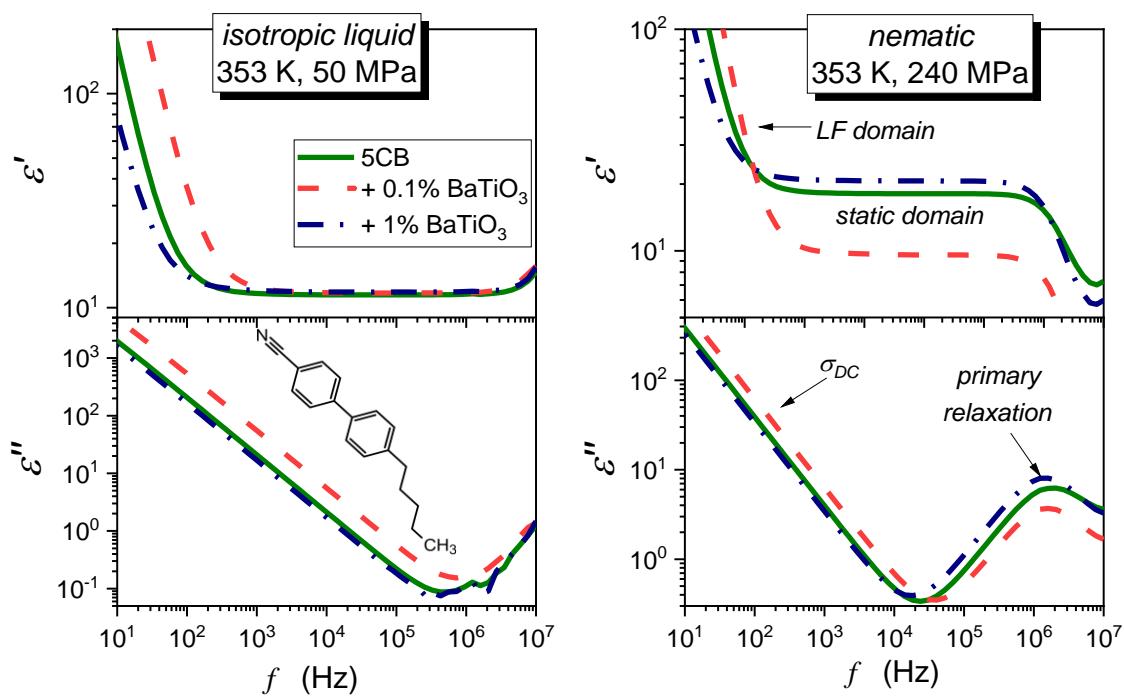

**Figure 1**  Examples of spectra obtained in BDS monitoring, presented via frequency evolutions of the real ($\varepsilon'$) and imaginary ($\varepsilon''$) parts of dielectric permittivity in tested 5CB and its nanocolloids under pressure Significant features and domains are indicated. They are the low-frequency (LF) domain, the static domain associated with dielectric constant, the primary loss curve, and the frequency part used for determining DC electric conductivity.

The body of the pressure chamber was surrounded by a jacket, enabling the circulation of a liquid from the large volume ($V = 20$ L) Julabo thermostat. The temperature was measured inside the chamber using copper – constantan thermocouple ($\pm 0.02$K) and two Pt100 resistors placed in the body of the chamber, to test possible temperature gradients. Changes in pressure were created via the computer-controlled pressure pump and measured using the tensometric pressure meter ($\pm 0.2$MPa). BDS studies were carried by means of Novocontrol Alpha-A analyzer in the frequency range from 1 Hz to 10 MHz. This range was associated with essential problems in BDS studies for higher frequencies. The voltage of the measuring field $U = 1$ V was applied. Examples of dielectric permittivity spectra for the tested compound are shown in Figure 1. It also presents significant features of such spectra and the structure of 5CB molecule. As the reference frequency for dielectric constant $\varepsilon = \varepsilon'(f = 116 kHz)$, in the mid of the static domain, was taken. The primary relaxation time was determined from the peak of the primary loss curve in the high-frequency part of $\varepsilon''(f)$ spectrum as $= 1/2\pi f_{peak}$, and DC electric conductivity from its low-frequency (LF) part: $\sigma = 2\pi f \varepsilon''(f)$ [61].

**III. Results and Discussion**

Figure 2 shows the pressure evolution of dielectric constant in 5CB and tested nanocolloids, covering isotropic (I), nematic (N), and solid (S) crystal phases. Figure 3 presents the insight focused on the evolution in the isotropic liquid phase. For the latter, all pretransitional effects are portrayed by the parallel of Eq. (4):

$$\varepsilon(P) = \varepsilon^* + a_P(P^* - P) + A_P(P^* - P)^{1-\alpha} \qquad , \quad T = const \qquad (5)$$

where $(P^*, \varepsilon^*)$ are for the extrapolated hypothetical continuous I-N transition, $P < P_{I-N} = P^* - \Delta P^*$, and $\Delta P^*$ is the pressure metric of the I-N phase transition discontinuity; $a_P$ and $A_P$ are constant amplitudes.

The similarity of the isobaric, temperature-related Eq. (4) and the isothermal pressure-related Eq. (5) with the same value of the critical exponent $\alpha$ agrees with the isomorphism postulate for critical phenomena [4], and earlier dielectric studies of pretransitional effects in critical liquids [17, 39, 57, 58, 62, 63].

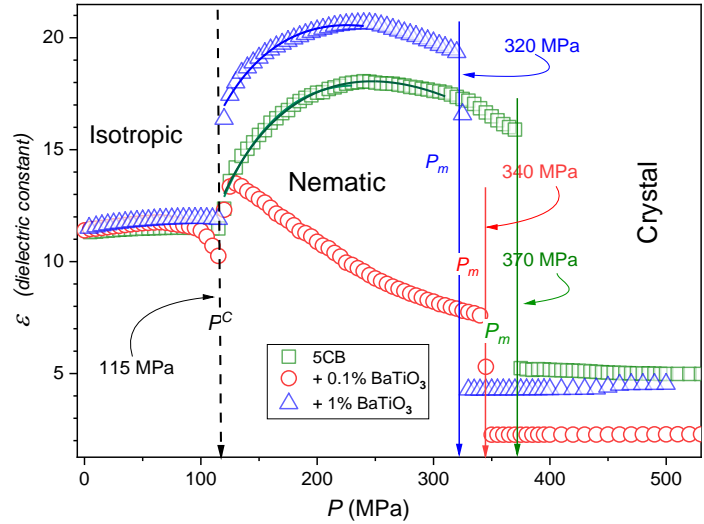

**Figure 2** Temperature evolutions of dielectric constant in compressed 5CB and its nanocolloids with BaTiO$_3$ nanoparticles on compressing, for $T = 353$ K isotherm.

The pretransitional anomaly described by Eq. (4) is associated with the crossover:

$$d\varepsilon(T)/dT < 0 \quad \to \quad d\varepsilon(T)/dT > 0 \quad \text{for } P = const \qquad (6)$$

It is associated with the prenematic fluctuations with permanent dipole moment ordered in an antiparallel way, leading to the cancellation of the dipolar component to dielectric constant. Consequently, the dielectric constant within fluctuations is significantly lesser than for the isotropic liquid surrounding. Near I-N transition the volume occupied by prenematic fluctuations dominates, yielding the above crossover. Similar patterns associated with the crossover to the fluctuations-dominated region near I-N transition occur on compressing in the isotropic phase, namely:

$$d\varepsilon(P)/dP > 0 \quad \to \quad d\varepsilon(P)/dP < 0 \quad \text{for } T = const \qquad (7)$$

Values of fitted parameters are given in Table I.

**Table I** Values of parameters describing pressure-related pretransitional changes of dielectric constant ((Eq. 1) and Fig. 3) in the isotropic phase of 5CB and its nanonocolloids.

| Parameter | 5CB | + 0.1% NP | + 1% NP |
|---|---|---|---|
| $\varepsilon_{I-N}$ | 11.48 | 10.25 | 11.90 |
| $P_{I-N}$ (MPa) | 115 | 115 | 115 |
| $a_P$ | -0.02 | -0.04 | -0.02 |
| $A_P$ | 0.32 | 0.54 | 0.25 |
| $\alpha$ | 0.5 | 0.5 | 0.5 |
| $\varepsilon^*$ | 9.82 | 9.72 | 11.04 |
| $P^*$ (MPa) | 197 | 115 | 147 |
| $\Delta P^*$ (MPa) | 82 | 0.2< | 32 |

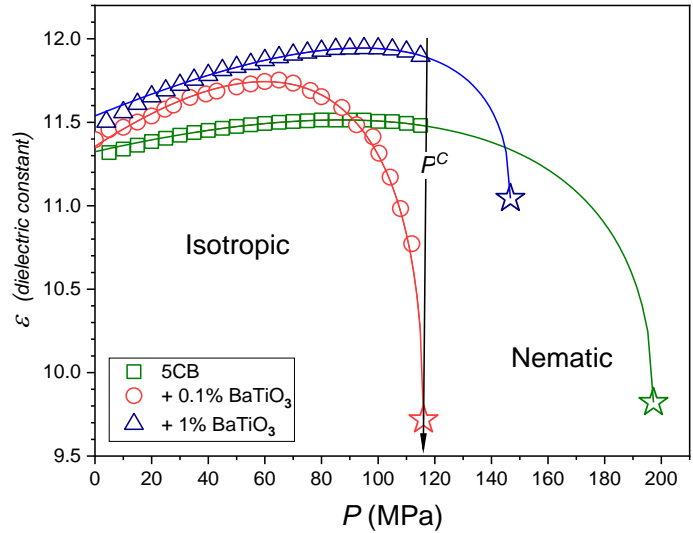

**Figure 3** The detailed insight into the pretransitional behavior of dielectric constant in the isotropic liquid phase of 5CB and its nanocolloids with $BaTiO_3$ nanoparticles for the pressure-related approach to the I-N transition. Solid curves are related to Eq. ( ) with parameters given in Table I. Stars denote the hypothetical extrapolated phase transition. The vertical arrow is for the I-N phase transition pressure detected in experiments.

The most striking feature of the results presented in Figs. 2 and 3 is the continuous I-N transition ($\Delta P^* \approx 0$) detected for a tiny addition ($x = 0.1\%$) of nanoparticles. When NPs increase concentration up to $x = 1\%$, the discontinuity is much larger than for the pure 5CB.

In the nematic mesophase, often samples 'oriented' for detecting parallel ($\varepsilon_\parallel$) and perpendicularly ($\varepsilon_\perp$) dielectric constant and electric polarizability components are studied, reflecting the rod-like or ellipsoidal symmetry of LC molecules. It can be realized via the extremely strong magnetic field, which is experimentally impossible for high-pressure dielectric studies. The alternative experimental approach is associated with covering capacitor plates with a polymeric agent supporting the required orientation of LC molecules.. However, pressure can distort the layer due to its sensitivity to compression compared to 'insensitive capacitor's plates made from Invar. It can resemble biofilm removal from a metal surface via compressing [64]. Such pressure action can also introduce parasitic impurities. Notable that phase transition in LC compounds is extremely sensitive to such contaminations [4, 65, 66]. Additionally, mentioned studies require a micrometric gap of the capacitor, introducing the restricted geometry to the system. This report is focused on bulk properties.

Despite the lack of any orientation-inducing treatments, samples of pure 5CB and its $x = 1\%$ nanocolloid follow the pattern associated with a strong rise of dielectric constant in the nematic phase on compressing, resembling the patterns which can be expected for $\varepsilon_\parallel(P)$ in the oriented sample. On compressing in the nematic phase dielectric constant increases up to $\varepsilon \approx 18.3$ in pure 5CB and $\varepsilon \approx 20.9$ in the nanocolloid, for $P \approx 233\ MPa$. These values are higher than ever observed for $\varepsilon_\parallel(T)$ in nematic 5CB. One can suppose that the extreme uniaxial ordering can result from the synergic impacts of compressing and BaTiO$_3$ nanoparticles influencing the orientation of the main part of 5CB molecules, directly associated with the permanent dipole moment. Notable that the allyl alkyl chain in n-cyanobiphenyls (nCB) molecules are slightly tilted to the direction of the main part of the molecule containing two phenyl rings. This causes that when using the magnetic field or electric field for the alignment of 5CB molecules in temperature studies under atmospheric pressure, the permanent dipole moment is inherently tilted with respect to the preferred direction, and only its component related to the projection of the dipole moment on the preferred direction determined by the external field is detected. Compressing can reduce this factor. In nanocolloids, adding some optimal amount of para-electric $BaTiO_3$ can reduce the steric hindrance, increasing the dipole moment contribution to polarizability and dielectric constant.

Both for pure 5CB and its $x=1\%$ nanocolloid dielectric constant significantly decreases when compressing for $P > 233 MPa$ in the nematic phase. There are no comparable pressure-related studies, but in available temperature-related studies, the constant values of $\varepsilon_\parallel$ and $\varepsilon_\perp$ are most often reported in the nematic phase. It is applied as the argument for the reliable determining of the anisotropy of dielectric constant $\Delta\varepsilon = \varepsilon_\parallel - \varepsilon_\perp$, one of the essential material characterization of rod-like LC materials. Notwithstanding, there is limited evidence showing $\varepsilon_\parallel(T)$ decrease in cooling, i.e., the 'bending behavior on cooling in the nematic phase. The origins of this phenomenon are not commented. Recently, the authors of this report also found the explicit manifestation of such a phenomenon in temperature studies of 5OCB and its nanocoloid, and commented it as a possible consequence of supercooling, , indeed occurring in recalled tests.

Such an explanation can also be implemented for the 'bending evolution' in the nematic phase presented in Fig. 2. Notwithstanding, we tested yet another hypothesis. Namely, the portrayal following the possible relation for $\varepsilon_\parallel(P)$ changes, which can be concluded from the known dependence for $\varepsilon_\parallel(T)$ changes [36, 37], namely:

$$\varepsilon_\parallel(P) = \varepsilon^{**} + B_P(P - P^{**})^\beta + A_P(P - P^{**})^{1-\alpha} + a_P(P - P^{**}) \Rightarrow \qquad (8)$$

For reducing the number of adjusting of parameters, we implemented the concept of the effective critical exponent in which the second power term is considered as a weighted correction to the first leading term [660]:

$$\varepsilon_\parallel(P) = \varepsilon^{**} + B_P(P - P^{**})^{\beta_{eff.}} + a_P(P - P^{**}) \qquad (9)$$

Figure 2 contains curves plotted on the base of Eq. (9) and fairly well portraying experimental data with the exponent $\beta_{eff.} \approx 0.65$ what can indicate the value $\beta = 1/2$ for the pressure paths in the nematic phase for pure 5CB and its $x=1\%$ nanocolloid. This result shows that the bending down remote from I-N transition is the inherent feature of the I-N pretransitional behavior in the nematic phase, which should be expected for a wide enough range of the nematic phase. This result can also ultimately explain and validate the 'bending' changes of dielectric constant observed in temperature studies.

For $x = 0.1\%$ NPs nanocolloid, dielectric constant in the nematic phase on compressing first slightly rises and subsequently permanently decreases up to the terminal freezing in the solid phase. In the authors' opinion, it can be associated with extreme sensitivity to external impacts, such as capacitor plates, characteristic of the continuous phase transition. For the solid phase dielectric constant of this sample decreases to $\varepsilon \approx 2.2$. Such value is typical for a complete crystalline phase with frozen orientational and translational freedom. For pure 5CB results presented in Fig. 2 show $\varepsilon \approx 4.3$, and for 5CB+0.1 BaTiO$_3$: $\varepsilon \approx 5.2$. Hence, the question arises for some influence of the plastic crystal phase in the given case, which may enlarge dielectric constant.

On decreasing frequency below the static domain (see Fig.1), the strong rise of $\varepsilon'(f)$ takes place. It is associated with the increasing impact of the ionic-related, translational contribution to polarizability. This low-frequency domain is extensively tested due to its particular significance for practical applications. Notwithstanding, the impact of pretransitional fluctuations on dielectric properties in the LF domain remains puzzling in pure LC and its nanocolloids. To the best of the authors' knowledge, only in refs. [67, 68] for 5CB and 11CB it was shown that the 'matrix' created by the pretransitional fluctuations leads to linear changes of the ionic contribution to dielectric permittivity, namely:

$$\Delta\varepsilon(f,T) = \varepsilon'(f,T)_{LF} - \varepsilon(T) = A(f) + B(f) \times T \tag{10}$$

where $\varepsilon(T)$ is for the reference dielectric constant change determined for $\varepsilon'(T)$ in the static domain, and "LF" indicates frequencies in the low-frequency region.

Figure 4 shows that similar behavior occurs for pressure-related behavior, namely

$$\Delta\varepsilon(f,P) = \varepsilon'(f,P)_{LF} - \varepsilon(T) = A(f) + B(f) \times P \quad , T = const \tag{11}$$

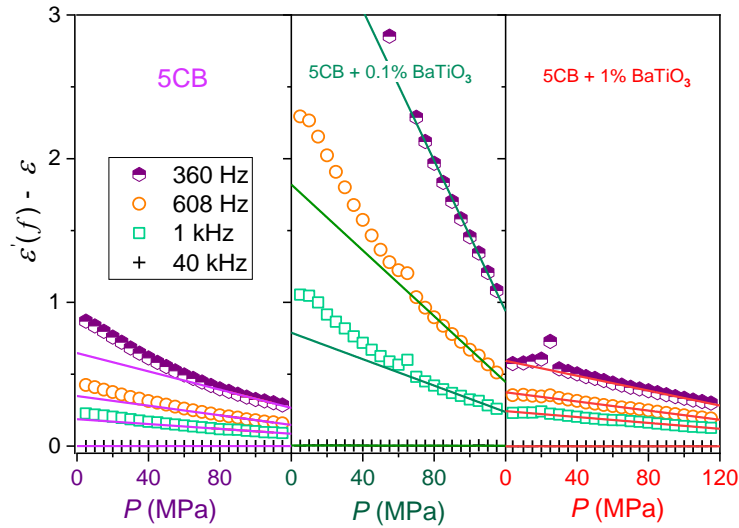

**Fig. 4** Pressure evolutions of the ionic contribution to dielectric constant (Eq. 11) in the isotropic liquid phase of 5CB and its tested nanocolloids.

Notable that the most considerable impact of the ionic contribution takes place for $x = 0.1\%$ BaTiO$_3$ nanocolloid, i.e., for the system showing the continuous I-N transition.

The limited frequency range in high-pressure studies enabled tests of basic properties characterizing dynamics only in the nematic phase. Figure 5 shows the primary relaxation time pressure changes in tested systems.

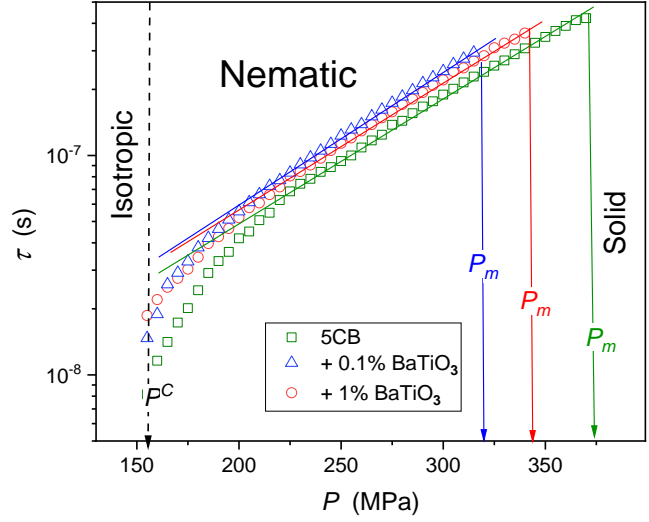

**Figure 5**     Changes of the primary relaxation time in the nematic phase of 5CB and its nanocolloids with BaTiO$_3$ nanoparticles on compressing for $T$ = 353 K isotherms. The applied scale facilitates showing the basic Barus relation, as linear dependences.

Generally, for pressure-related changes in relaxation time, one can expect the description via the super-Barus (SB) relation [57, 58]:

$$\tau(P) = \tau_0 exp\left(\frac{PV_a(P)}{RT}\right) \tag{12}$$

where $T = const$, $V_a(P)$ is the apparent activation volume, $R$ denotes the gas constant.

As visible in Fig. 5, remote from I-N transition, the basic Barus relation with $V_a(P) = V_a = const$ emerges. It is associated with the same activation volume $V_a = 40\ cm^3/mol$ for each tested system. Strong, pretransitional distortions occur near I-N transition, starting from $P_{I-N} - 60 MPa$. The impact of nanoparticles is here different than for dielectric constant. Namely $\tau(P)$ changes in pure 5CB (discontinuous transition) and 5CB+0.1% BaTiO$_3$ nanocolloids (continuous phase transition) follow the same track. Notable that the primary relaxation process is notably faster in pure 5CB than in related nanocolloids.

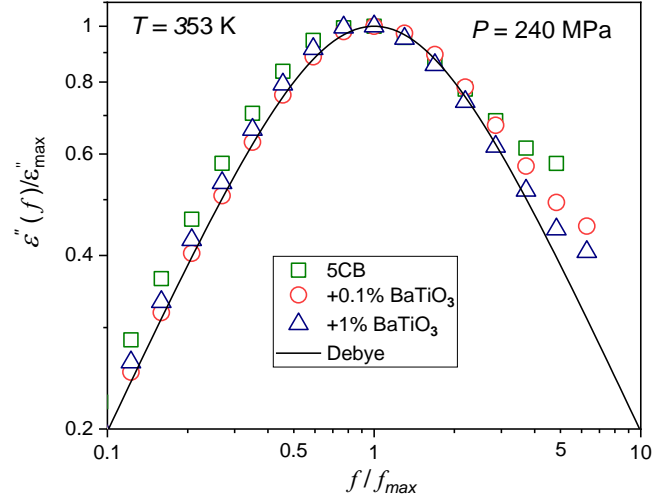

**Figure 6**  The normalized superposition of dielectric loss curves in the mid of the nematic phase reached due to compressing for 5CB and its nanocolloids with BaTiO$_3$ nanoparticles.

Figure 6 shows the Jonsher-type [69] normalized superposition of dielectric loss curves for 5CB and tested nanocolloids, for the pressure in the mid of the nematic phase:

$$\varepsilon''(f < f_{peak}) \propto \left(\frac{f}{f_{peak}}\right)^m \quad , \quad \varepsilon''(f > f_{peak}) \propto \left(\frac{f}{f_{peak}}\right)^{-n} \tag{13}$$

For comparison, the Debye curve [61] related to the single relaxation time and $m = n = 1$ [69] is also plotted. For pure 5CB the broadening of the distributions of relaxation times, in comparison to the reference Debye curve, is visible, both for the low- ($f < f_{max}$) and high- ($f > f_{max}$) frequency parts. The addition of nanoparticles shifts the distribution towards the basic Debye pattern.

Studies of the evolution of the primary relaxation time are associated with the dynamics of orientational processes. Tests of DC electric conductivity contain message related to the translational processes. In complex systems, both properties are often decoupled, which is expressed by the fractional Debye-Stokes-Einstein (fDSE) relation [70]. Figure 6 shows its implementation for the pressure path, namely:

$$\sigma(P)[\tau(P)]^S = C = const \quad , \qquad T = const \tag{13}$$

where the exponent $S = 1$ is related to the translational-orientational coupling, i.e., to the basic reference DSE law.

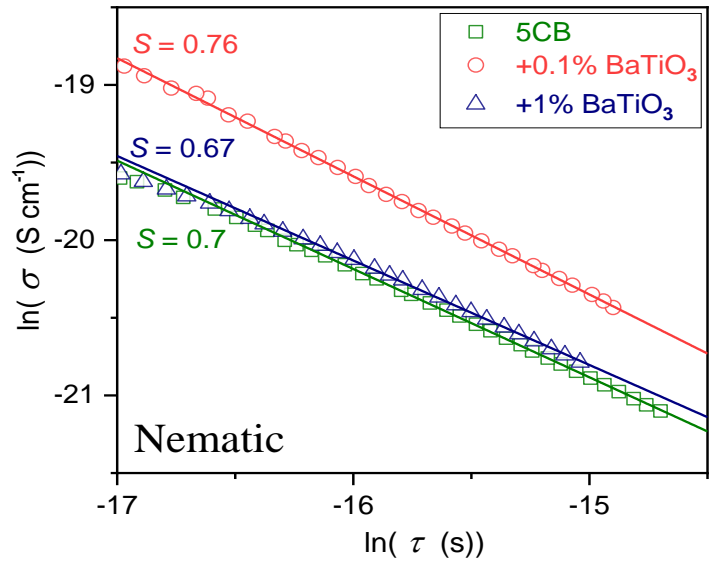

**Figure 7** Test of the translational-orientational decoupling in the nematic phase of compressed 5CB and its nanocolloids with BaTiO$_3$.

Figure 7 presents the explicit fDSE behavior in tested systems. Pure 5CB and its $x$=1% nanocolloid in the nematic phase show a similar degree of decoupling. It is notably lesser for $x$=0.1% nanocolloid. Significant hallmarks of phase transitions and pretransitional effects strongly manifested in temperature studies under atmospheric pressure are very limited in the reported pressure test.

Furthermore, the weakening of the I-N phase transition for small concentrations of BaTiO$_3$ could be several reasons. The interlayer correlation can be reduced for a small concentration of NPs. In this case, the coupling between the orientational order parameter and nanoparticle order parameter decreases, resulting in such a weakening of the transition. For a small concentration of the paraelectric BaTiO$_3$ reduces the pressure range of the isotropic phase. The pressure range of hysteresis associated with the I-N transition becomes smaller when the pressure range of the isotropic phase is reduced. Consequently, the I-N transition, which has a first-order character in bulk, changes over to a continuous one for small concentration of BaTiO$_3$ NPs. Owing to the restricted size of the nanoparticle to which the correlations can grow, the disorder is substantial at a small concentration of BaTiO$_3$. Therefore, at a small concentration of NPs, the nematic phase occurs at a much higher and exists over a larger range of pressure. The presence of disorder weakens the strength of the transition, causing the first-order character of the I-N transition to change to the continuous one

**Conclusions**

The report shows the first comprehensive test of static and dynamic dielectric properties in compressed rod-like nematogenic liquid crystalline compound (5CB) and its nanocolloids with BaTiO$_3$ nanoparticles. Studies extend from the ionic low-frequency domain to the relaxation region, focusing on the impact of pretransitional fluctuations.

For the nanocolloid related to $x$=0.1% of BaTiO$_3$ nanoparticles the continuous, in the limit of the experimental error, I-N transition has been detected. In commenting on this unique result, we would like to recall the conclusion of ref. [71], developed originally for the nematic transition in elastomeric soft materials ' … *weak random anisotropy results in a singular renormalization of the Landau-De Gennes expression, adding an energy term proportional to the inverse quartic power of order parameter Q. This reduces the first-order discontinuity … For sufficiently high disorder strength the jump disappears altogether and the phase transition becomes continuous, in some ways resembling the supercritical transitions in external field …*' In this report, the mentioned disorder can introduce the proper concentration of BaTiO$_3$ nanoparticles, which impact is additionally tuned by compressing.

Notable is also the possibility of creating the 'endogenic orientation' by compressing matched with the addition of nanoparticles, leading to the enormous increase of dielectric constant. Regarding this issue is worth indicating the new explanation for the 'bending' behavior of dielectric constant observed earlier in nematic test under atmospheric pressure for systems with a wide range of the nematic phase. The analysis in the given report links it to long-range consequences of the pretransitional behavior.

The synergic impact of nanoparticles and pressure on the nematic phase also led to the large 'jump' of dielectric constant at the nematic – solid (N-S) strongly discontinuous transition. Following refs. [72, 73] such a change of dielectric constant can introduce a remarkable additional contribution to the phase transition entropy $\Delta S_E \sim \Delta \varepsilon / T \sim 60 J K^{-1} kg^{-1}$ supplementing the already large reference value recalled in the Introduction. Hence one can expect the Collossal barocaloric effect [74, 75]: in the given case 5CB and 5CB nanocolloids systems, already at very moderate pressures.

Notable is also the glassy dynamics detected in nematic 5CB and its nanocolloids, although the impact of pretransitional fluctuations under compression seems to be lesser than in reference temperature tests under atmospheric pressure.

**Declaration of Competing Interest**

The authors declare that they have no known competing financial interests or personal relationships that could have appeared to influence the work reported in this paper.

**Authors Contributions**

JŁ was responsible for measurements, data analysis, and paper preparations, ADR supported data analysis and paper preparation, SJR supported paper preparation.

**Experimental data availability:**

The author declares the availability of experimental data upon reasonable request.

**Authors reference data:**

Joanna Łoś: ORCID: 0000-0001-6283-0948 ; e-mail: joalos@unipress.waw.pl

Aleksandra Drozd-Rzoska: ORCID: 0000-0001-8510-2388 ;

e-mail: Ola.DrozdRzoska@gmail.com

Sylwester J. Rzoska: ORCID: 0000-0002-2736-2891 ; e-mail: sylwester.rzoska@gmail.com

Szymon Starzonek: ORCID: 0000-0003-2793-7971 ; e-mail: starzonek@icloud.com

Krzysztof Czupryński: ORCID: 0000-0002-1785-7786 ; e-mail: krzysztof.czuprynski@wat.edu.pl

Prabir K. Mukherjee: ORCID: 0000-0003-4523-5391; e-mail: pkmuk1966@gmail.com